\documentclass{article}

\usepackage{amsfonts}

\def\be{\begin{equation}}
\def\ee{\end{equation}}
\def\bea{\begin{eqnarray}}
\def\eea{\end{eqnarray}}
\def\({\left(}
\def\){\right)}
\def\<{\left<}
\def\>{\right>}

\def\[{\left[}
\def\]{\right]}
\def\tr{{\mbox{tr}}}
\def\be{\begin{equation}}
\def\ee{\end{equation}}
\def\bea{\begin{eqnarray}}
\def\eea{\end{eqnarray}}
\def\({\left(}
\def\){\right)}
\def\<{\left<}
\def\>{\right>}

\def\[{\left[}
\def\]{\right]}

\def\+{\bar}
\def\mb{\mathbb}
\def\tr{{\mbox{tr}}}

\def\A{{\cal{A}}}
\def\F{{\cal{F}}}
\def\U{{\cal{U}}}
\def\a{{\cal{A}}}
\def\b{{\cal{B}}}
\def\c{{\cal{C}}}

\def\rep{\mbox{rep}}

\begin{document}

\pagestyle{empty}
\vskip-10pt
\vskip-10pt
\hfill 

\begin{center}
\vskip 3truecm
{\Large \bf
Closed non-abelian strings}\\ 
\vskip 2truecm
{\large \bf
Andreas Gustavsson}\footnote{a.r.gustavsson@swipnet.se}\\
\vskip 1truecm
{\it F\"{o}rstamajgatan 24,\\
S-415 10 
G\"{o}teborg, Sweden}\\
\end{center}

\vskip 2truecm
{\abstract{With the aim of finding a framework for describing $(2,0)$ theory, we propose a non-abelian gerbe with surface holonomies that can parallel transport closed strings only.}}
\vfill 
\vskip4pt
\eject
\pagestyle{plain}

\section{Introduction}
The appropriate framework for describing $(2,0)$ supersymmetric gauge theories in six dimensions should be some non-abelian gerbe \cite{Witten}.

Non-abelian gerbes with holonomy\footnote{By holonomy we will in this Letter always mean the operator that parallel transports a charged object along a given path, and this path need not be closed.} have been constructed in \cite{Schreiber} and references therein. In these references, a connection one-form on loop space is constructed in terms of a connection one-form and a connection two-form in space-time whereof the one-form has to be subject to a certain flatness constraint (vanishing fake curvature) thus to be imposed in space-time rather than in loop space.

It is not known how, nor if, this construction can be applied to $(2,0)$ theory. The action for these connection one-forms and two-forms is not known. Complications seems to arise because of the flatness constraint that has to be imposed. It does not seem to be clearly understood how this constraint should arise from some $(2,0)$ supersymmetric action (possibly using some auxiliary Lagrange multiplier field).

Things might get more clear in loop space. We think the non-abelian gerbe on space-time, to be used to describe $(2,0)$ theory, should correspond in loop space to a quite ordinary fiber bundle, albeit over an infinite-dimensional loop space. And that the (classical) action should be a quite ordinary Yang-Mills action on that loop space.

We can not use the one-form constructed in \cite{Schreiber} for this purpose though, since that construction relied on a flatness condition imposed on a connection one-form in {\sl{space-time}}. In particular it would be insufficient to just study the Yang-Mills action in loop space. We would also have to find a way to get the flatness constraint from some action in loop space. That would require us to find a way to formulate the flatness constraint in terms of loop space fields only. But that does not seem to be possible. This is the main reason why I would like to suggest an alternative construction of a non-abelian gerbe, that could enable us to formulate the action principle entirely in loop space. 

We also note that the construction in \cite{Schreiber} is based on the assumption that surface holonomies parallel transport {\sl{open}} strings.\footnote{From this it should of course be possible to derive a holonomy that parallel transports closed strings, simply by gluing together the end points of two open strings to form a closed string.} This assumption leads to restrictions on the surface holonomies that would be absent if we only had closed strings at our disposal. The restrictions come from the many ways of composing surface holonomies associated with open strings: We can glue together two surfaces (both of the topology of a disk) along a common open boundary string (this is what may be called vertical composition of surface elements if one draws the common boundary string horizontally). We can also attach an open string at the end point of another open string (horizontal composition). Requiring that vertical and horizontal compositions of surface holonomies commute, one finds that only abelian gauge groups can be allowed, unless one introduces line holonomies as well. This connection one-form associated with these line holonomies must then be subject to the flatness constraint. 

Any object that can be parallel transported by means of some connection, should in physics correspond to a charged physical object that couples electrically to that connection. Given that the boundary of an $M2$ brane that ends on a stack of $M5$ branes is a closed string\footnote{or an infinitely extended string -- in any case a string without a boundary}, one may wonder where the open strings and the point particles could come from. 

General principles, like the gauge principle, the action principle, supersymmetry, etcetera, are likely to be best understood when formulated in loop space. We will define loop space over a manifold $M$ as in \cite{Hitchin}, that is, as the space of mappings from $S^1$ into $M$. An abelian gerbe on $M$ is a line bundle on $LM$, and a surface holonomy is an ordinary Wilson line in $LM$. In loop space no open strings on $M$ can be parallel transported by holonomies. Only closed strings can.

With these motivations, we will assume that only closed strings can be parallel transported, which thus lead us to what seems to be an alternative way of defining non-abelian gerbes, entirely in loop space. 

We begin, in section \ref{abel}, by introducing the abelian gerbe with a connection two-form, using the language of Cech \cite{Hitchin}, and show in section \ref{loop} how this  leads to a line bundle in free loop space \cite{Hitchin}. In particular we obtain an explicit form for the connection one-form in loop space as the transgression of the connection two-fom. To my knowledge this way of expressing the one-form on loop space as the transgression of a local two-form connection over many charts, has not appeared in the literature before. (If one wants a loop to live in just one chart, one would in general have to use charts that depend on the loop, which again would involve a mixture of loop space and space-time concepts that I do not find very appealing). In section \ref{nonabel} we generalize the loop space picture of the gerbe to non-abelian gauge groups.

\section{Abelian gerbes}\label{abel}
In this section we review the concept of a connection on an abelian gerbe using the Cech language \cite{Hitchin}. Given a manifold $M$, which we can cover with open charts $U^{\alpha}$ in such a way that each chart and each overlap is diffeomorphic to $\mb{R}^m$, an abelian gerbe is specified by its transition functions $g^{\alpha\beta\gamma}$ on triple overlaps $U^{\alpha\beta\gamma}:=U^{\alpha}\cap U^{\beta}\cap U^{\gamma}$. These take values in the $U(1)$ gauge group,
\bea
g^{\alpha\beta\gamma}=e^{if^{\alpha\beta\gamma}}.
\eea
The $f^{\alpha\beta\gamma}$ is completely antisymmetric in the Cech indices $\alpha,\beta,\gamma$. Following \cite{Alvarez} we will assume the same antisymmetry of the overlaps, so that for instance $U^{\alpha\beta}=-U^{\beta\alpha}$ where minus sign here means orientation reversal. The Cech coboundary operator $\delta$ acts as 
\bea
(\delta f)^{\alpha\beta\gamma\delta} := f^{\beta\gamma\delta} - f^{\alpha\gamma\delta} + f^{\alpha\beta\delta} - f^{\alpha\beta\gamma}
\eea
on $U^{\alpha\beta\gamma\delta}$, and with the obvious generalization to any number of intersections. The transition functions are subject to the cocycle condition (here with multiplication in place of addition)
\bea
(\delta g)^{\alpha\beta\gamma\delta} = 1
\eea
on quadruple overlaps $U^{\alpha\beta\gamma\delta}$. That means that 
\bea
(\delta f)^{\alpha\beta\gamma\delta} \equiv 0 {\mbox{ modulo $2\pi$.}}\label{coc}
\eea

A connection on an abelian gerbe is a collection of locally defined two-forms $B^{\alpha}$ defined on $U^{\alpha}$. On double overlaps these are subject to the patching conditions
\bea
B^{\alpha}-B^{\beta} = d\Lambda^{\alpha\beta}.
\eea
where $\Lambda^{\alpha\beta}$ are some one-forms defined on $U^{\alpha\beta}$. On triple overlaps they are subject to 
\bea
(\delta\Lambda)^{\alpha\beta\gamma}=df^{\alpha\beta\gamma}.
\eea
Due to $\delta\delta=0$, we have some gauge freedom in our choice of transition functions, and the other data that specifies the connection,
\bea
f^{\alpha\beta\gamma} &\rightarrow & f^{\alpha\beta\gamma} + (\delta h)^{\alpha\beta\gamma}\cr
\Lambda^{\alpha\beta} &\rightarrow & \Lambda^{\alpha\beta} + dh^{\alpha\beta} + (\delta\lambda)^{\alpha\beta}\cr
B^{\alpha} &\rightarrow & B^{\alpha}+d\lambda^{\alpha}.
\eea
This freedom is a consequence of repeated use of the Poincare lemma.

\subsection{The Wilson surface}
Given such a two-form connection $B$, a Wilson surface associated with a surface $\Sigma$, is defined, in the spirit of \cite{Alvarez}, as
\bea
W(\Sigma)=\exp {i\int_{\Sigma} B}
\eea
where 
\bea
{\int_{\Sigma} B} := \sum_{\alpha} \int_{V^{\alpha}} B^{\alpha} - \sum_{\alpha,\beta} \int_{V^{\alpha\beta}} \Lambda^{\alpha\beta} + \sum_{\alpha,\beta,\gamma} \int_{V^{\alpha\beta\gamma}} f^{\alpha\beta\gamma}.\label{expon}
\eea
Here the charts $V^{\alpha}\subset U^{\alpha}\cap \Sigma$ are maximally contracted, by which we mean that double overlaps between such charts are of one dimension lower. An example of maximally contracted charts is a triangulation of $\Sigma$. The $V^{\alpha\beta}=V^{\alpha}\cap V^{\beta}$ are of one dimension lower than the $V^{\alpha}$, and the $V^{\alpha\beta\gamma}$ of one dimension lower than $V^{\alpha\beta}$. By integration over a point we mean evaluation at that point. 

Using techniques in \cite{Alvarez}, it can be shown that Eq (\ref{expon}) can only change by some $\delta f \equiv 0$ modulo $2\pi$ if one would modify the open covering. Hence the exponent is well-defined modulo gauge variations.

Under a gauge variation\footnote{Here we use $\delta_v$ to denote variation, to distinguish variation from Cech coboundary.} 
\bea
\delta_v B^{\alpha} &=& d\lambda^{\alpha}\cr
\delta_v \Lambda^{\alpha\beta} &=& dh^{\alpha\beta} + \lambda^{\beta}-\lambda^{\alpha}\cr
\delta_v f^{\alpha\beta\gamma} &=& h^{\beta\gamma} - h^{\alpha\gamma} + h^{\alpha\beta} \label{gaugevariation}
\eea
we find that
\bea
\delta_v {\int_{\Sigma} B} = \int_C \lambda
\eea
where we define
\bea
\int_C \lambda:=\sum_{\alpha} \int_{C^{\alpha}} \lambda^{\alpha} - \sum_{\alpha,\beta} \int_{C^{\alpha\beta}} h^{\alpha\beta}
\eea
Here $C^{\alpha}$ is the piece of the boundary $\partial V^{\alpha}$ which is not adjacent to any other boundary $\partial V^{\beta}$, that is, it is a boundary of $\Sigma$. If $\Sigma$ is closed, we of course do not have any such boundary $C$ and hence the closed Wilson surface is gauge invariant.

We can express the change of the Wilson surface as
\bea
W(\Sigma)&\rightarrow& g(C)W(\Sigma).
\eea
where $\partial \Sigma = C$ may be one or several disjoint boundary loops, and
\bea
g(C) := \exp i\int_C \lambda.
\eea

\subsection{Magnetic charge}
The curvature $H=dB$ is a globally defined three-form that defines an element in $H^3(M,2\pi\mb{Z})$, which means that 
\bea
\int_{M_3} \frac{H}{2\pi} = \frac{1}{2\pi}\sum_{\alpha,\beta,\gamma,\delta} (\delta f)^{\alpha\beta\gamma\delta} 
\eea
is an integer. Here the sum is over points where four maximally contracted charts intersect. \footnote{If $V_{\alpha}$ is a three-dimensional maximally contracted chart, then $V_{\alpha\beta}$ is two-dimensional, $V_{\alpha\beta\gamma}$ is one-dimensional and $V_{\alpha\beta\gamma\delta}$ zero-dimensional.} The short computation required to show this identity is done in the spirit of \cite{Alvarez}.

\section{Loop space}\label{loop}
Free loop space $LM$ is given by all mapping Map$(S^1,M)$. In other words, it is the space of parametrized loops $C: s\mapsto C^{\mu}(s)$ in $M$. We use as coordinates in $LM$, 
\bea
C^{\mu}(s)
\eea
where $s$ is to interpreted as a continuous index. We have tangent vectors given by the functional derivatives
\bea
\partial_{\mu s} := \frac{\delta}{\delta C^{\mu}(s)}
\eea
that span the tangent space $T_{C}LM$, and we have associated co-tangent vectors
\bea
\delta C^{\mu}(s)
\eea
It may sometimes be convenient to form quantity that has all the index up-stairs
\bea
dC^{\mu s} := \frac{ds}{2\pi} \delta C^{\mu}(s).
\eea
The exterior derivative is now given as
\bea
\delta = \int \frac{ds}{2\pi} \delta C^{\mu}(s)\partial_{\mu s} := dC^{\mu s}\partial_{\mu s}
\eea
which is consistent with the notation as
\bea
\delta C^{\mu}(s) = \int \frac{dt}{2\pi} \delta C^{\nu}(t)\partial_{\nu t} C^{\mu}(s)
\eea
With this measure we thus have
\bea
\partial_{\mu s}C^{\nu}(t) = 2\pi \delta_s(t) \delta^{\nu}_{\mu}.
\eea
where we denote the Dirac delta function supported at $s$ as $\delta_s(t):=\delta(s-t)=\delta(t-s)=\delta_t(s)$, just to display its covariance property. The identity map is given by
\bea
\delta^s_t = ds \delta_t(s)
\eea
and has the properties
\bea
V_s\delta^s_t &=& V_t\cr
U^t\delta^s_t &=& U^s
\eea
for any quantities $V_s$ and $U^s:= ds U(s)$.

Henceforth we will denote the exterior derivative by $d$ to avoid confusion with the Cech coboundary operator $\delta$ and ordinary variations.\footnote{It could be confusing to write $dC^{\mu}(s)$ in place of $\delta C^{\mu}(s)$ though, as the former could be interpreted as $ds \dot{C}^{\mu}(s)$, which is certainly not what is intended here. So I stick to the notation $\delta C^{\mu}(s)$ for the cotangent vectors.}

\subsection{Covariance in loop space}
We define a covariant vector $V_{\mu s}$ and a contravariant vector $U^{\mu s}$ in loop space as quantities that varies according to
\bea
\delta V_{\mu s} &=& \epsilon^{\rho r}\partial_{\rho r}V_{\mu s} + (\partial_{\mu s}\epsilon^{\rho r}) V_{\rho r}\label{vec}\\
\delta U^{\mu s} &=& \epsilon^{\rho r}\partial_{\rho r}U^{\mu s} - (\partial_{\rho r}\epsilon^{\mu s})U^{\rho r}\label{covec}
\eea
respectively, under an infinitesimal variation\footnote{Here $\delta$ denotes a variation, not the exterior derivative.}
\bea
\delta C^{\mu}(s) = -\epsilon^{\mu}(s;C)
\eea
in loop space. We define 
\bea
\epsilon^{\mu s} := \frac{ds}{2\pi} \epsilon^{\mu}(s)
\eea
and index contraction of $s$ means integration.

Given a two-form $B_{\mu\nu}$ in space-time, we can construct an associated covariant vector in loop space as
\bea
A_{\mu s} = B_{\mu\nu}(C(s))\dot{C}^{\nu}(s)
\eea
One may check\footnote{Let's do it here. Under a space-time diffeomorphism the two-form $B$ varies according to
\bea
\delta B_{\mu\nu} = \epsilon^{\rho}\partial_{\rho} B_{\mu\nu} + (\partial_{\mu}\epsilon^{\rho})B_{\rho\nu}+(\partial_{\nu}\epsilon^{\rho})B_{\mu\rho}
\eea
and this variation of $B$ induces the following variation of $A_{\mu s}$,
\bea
\delta A_{\mu s} &=& \epsilon^{\rho}\partial_{\rho}A_{\mu} + \dot{C}^{\nu}\partial_{\nu}\epsilon^{\rho} B_{\mu\rho} + (\partial_{\mu}\epsilon^{\rho})A_{\rho}\cr
&=& \epsilon^{\rho r}\partial_{\rho r}A_{\mu s} + (\partial_{\mu s}\epsilon^{\rho r}) A_{\rho r}.
\eea
In the last step we have noted that
\bea
\epsilon^{\rho r}\partial_{\rho r}A_{\mu s} &=& \int dr \epsilon^{\rho}(C(r)) \partial_{\rho r}\(B_{\mu\nu}(C(s))\dot{C}^{\nu}(s)\)\cr
&=& \epsilon^{\rho}\partial_{\rho}A_{\mu s} + \dot{C}^{\nu}(s) \partial_{\nu}\epsilon^{\rho}(C(s)) B_{\mu\rho}(C(s)).
\eea} that this quantity indeed transforms as a vector under loop space diffeomorphisms with parameters
\bea
\epsilon^{\mu}(s;C) = \epsilon^{\mu}(C(s))
\eea
induced by space-time diffeomorphisms
\bea
\delta x^{\mu} = -\epsilon^{\mu}(x).
\eea

We may also consider reparametrizations $\delta s=-\epsilon^s$ of the loops. Here the index $s$ shall be viewed as a vector index in one dimensions, and not as a continuous index. In loop space we would then consider the diffeomorphism with parameter
\bea
\epsilon^{\mu}(s;C)=-\epsilon^s\dot{C}^{\mu}(s)\label{rep}
\eea
which is to say that the $C^{\mu}(s)$ transform as scalars under a reparametrization. However in loop space we keep $C$ fixed, for instance when we consider the variation of $V_{\mu s}$,
\bea
\delta V_{\mu s}(C)\equiv V'_{\mu s}(C)-V_{\mu s}(C). 
\eea
Inserting (\ref{rep}) into (\ref{vec}) we find that
\bea
\delta V_{\mu s} = \epsilon^{\rho r}\partial_{\rho r} V_{\mu s} + \partial_s\(\epsilon^s V_{\mu s}\)
\eea
This differs from the variation under a reparametrization for which one keeps $s$ fixed (and let $C$ vary according to (\ref{rep})), 
\bea
\delta_{\rep} V_{\mu s}(C) \equiv V'_{\mu s}(C')-V_{\mu s}(C)
\eea
but we can relate these variations to each other as
\bea
\delta V_{\mu s}(C) &=& \delta_{\rep} V_{\mu s}(C) - \(V'_{\mu s}(C') - V'_{\mu s}(C)\)\cr 
&=& \delta_{\rep} V_{\mu s}(C) + \epsilon^{\rho r}\partial_{\rho r} V_{\mu s}(C).
\eea
and hence get
\bea
\delta_{\rep} V_{\mu s} = \epsilon^s\partial_s V_{\mu s} + (\partial_s\epsilon^s)V_{\mu s}
\eea
which means that $V_{\mu s}$ is a vector under reparametrizations.

Let us now study how $U^{\mu s}(C):=ds U^{\mu}(s;C)$ varies under a reparametrization. From Eq (\ref{covec}) we find that
\bea
\delta U^{\mu}(s) = \epsilon^{\rho r}\partial_{\rho r}U^{\mu}(s) + \epsilon^s\partial_s U^{\mu}(s)
\eea
and hence
\bea
\delta_{\rep} U^{\mu}(s) = \epsilon^s\partial_s U^{\mu}(s).
\eea
so this is a scalar transforms under a reparametrization. Now this is exactly what we need in order to build a scalar field on loop space as the contraction
\bea
S=V_{\mu s}U^{\mu s}
\eea
If now $V_{\mu s}$ is a covariant vector and $U^{\mu}(s)$ a scalar under reparametrizations, then we find that $S$ is reparametrization invariant,
\bea
\delta_{\rep} S = \int \frac{ds}{2\pi} \partial_s\(\epsilon(s)V_{\mu s}U^{\mu}(s)\) = 0.
\eea

\subsection{The connection in loop space}
The Wilson surface assocated with a surface $\Sigma$ with the topology of a cylinder with boundary loops $C_1$ and $C_2$ becomes a Wilson line $\Gamma$ between the points $C_1$ and $C_2$ in $LM$. In this picture we will view a Wilson surface with the topology of a disk as a degenerate cylinder where one of the boundary loops say $C_1$ has shrunk to a point $x$, and hence we view also the disk as a line in loop space -- a line from $x$ to $C_2$. A surface $\Sigma$ with an arbitrary number of boundary loops can also be viewed in loop space as a line $\Gamma$. We can for instance interpret all the disjoint loops as just one loop $C_1$, and let some point inside the surface be the second loop $C_2$. The line is then between the points $C_1$ and $C_2$. 

We can cover $M$ with open charts. But these open charts do not seem to be well suited for loops. While we can always find a chart such that a point belongs to it, this need not be the case for a loop. It can well happen that no chart contains that entire loop. The loop may well go over several charts. So it seems that we have to extend the concept of an open covering of $M$ to an open covering of $LM$ because a loop in $M$ is indeed a point in $LM$. The price we have to pay for this is that we get open charts in an infinite-dimensional loop space. Clearly the set of open charts in $LM$ is much bigger than the set of loop spaces over open charts in $M$. We will denote the open charts in $LM$ as $\U^{\a}$ and these have in general nothing to do with the open charts $U^{\alpha}$ in $M$. We may for instance notice that an open chart in $LM$ need not be topologically trivial in $M$. For instance a cylinder in $M$ is a line in $LM$. We can also give an example that shows that not all loop spaces are topologically trivial and hence must be covered by several charts. If $M$ is a two-torus then $LM$ contains a non-contractible one-cycle, and hence it must be topologically non-trivial. 

Given a fibration of our six-dimensional space, $\Pi: M \rightarrow M_5$, we can define open charts in $LM$ by means of open charts $U^{\a}$ on $M_5$ as follows. We define $LM_{\Pi,M_5}$ as the sector of $LM$ that consists of loops that are fibers in the bundle $\Pi: M \rightarrow M_5$. 

Now, can any loop be viewed as some fiber of some bundle? For instance, on a cylinder $R\times S^1$ it appears that only those loops which wind the non-trivial cycle $S^1$ may be interpreted as fibers in a bundle with base-manifold $R$. Any other fibration, in which the fibres do not wind the $S^1$, will inevitably involve some loop that shrinks to zero size. So only if we allow for some point-like loops in our fibrations, our construction can work in the full generality.

Then we can express almost all open charts in $LM$ as the union
\bea
\U^{\a}_{M_5} = \bigcup_{\Pi}\{C\in LM_{\Pi,M_5} | \Pi(C) \in U^{\a}\}
\eea
where $\Pi$ runs over all possible fibrations of $M$ over the base manifold $M_5$. This excludes the set of loops (which has zero measure in $LM$) which intersect $M_5$ at many points, or which lie within $M_5$. These loops must of course also belong to some open chart in $LM$. We can incorporate them in open charts that we define by using another base-manifold with other open charts. The total set of open charts in $LM$ will then be the union of open charts associated with each choice of base-manifold $M_5$. The total set of open charts in $LM$ is thus given by the union 
\bea
\bigcup_{M_5} \left\{\U_{M_5}^{\a}\right\}_{\a\in M_5}
\eea
where $\a$ runs over the open charts in each $M_5$ respectively.

We will use a Cech index notation such that 
\bea
\U^{\a} \subset \bigcup_{\alpha\in \a} U^{\alpha}
\eea
when both sides are seen as subspaces of $M$. 

Given a connection two-form $B$ on $M$, we obtain a connection one-form $\A$ on $LM$ as the transgression of $B$ over loops in $M$,
\bea
\A^{\a} = \sum_{\alpha\in \a} \int_{C_{\alpha}} B^{\alpha} - \sum_{\alpha,\beta\in \a} \Lambda^{\alpha\beta}
\eea
or, spelling it out in all details,
\bea
\A^{\a}(C) &=& \sum_{\alpha}\int_{C_{\alpha}} \frac{ds}{2\pi} B^{\alpha}_{\mu\nu}(C(s))\dot{C}^{\nu}(s)\delta C^{\mu}(s)\cr
&& - \sum_{\alpha\beta} \Lambda^{\alpha\beta}_{\mu}(C_{\alpha\beta})\delta C^{\mu}(s).\label{transgress}
\eea
On a double overlap $\U^{\a\a'}:=\U^{\a}\cap \U^{\a'}$, we have 
\bea
\A^{\a}-\A^{\a'} = d\Lambda^{\a\a'}
\eea
where \footnote{We use the convention that $\alpha\in \a$, $\alpha'\in \a'$ and so on.}
\bea
\Lambda^{\a\a'} = \sum_{\alpha,\alpha'} \int_{C_{\alpha}\cap C_{\alpha'}} \Lambda^{\alpha\alpha'} - \sum_{\alpha,\beta,\alpha'} f^{\alpha\beta\alpha'} - \sum_{\alpha,\alpha',\beta'} f^{\alpha\alpha'\beta'}.
\eea
To see this, we compute
\bea
d\Lambda^{\a\a'} &=& \sum_{\alpha,\alpha'} \int_{C_{\alpha\alpha'}} d\Lambda^{\alpha\alpha'} -  \sum_{\alpha,\alpha'} \Lambda^{\alpha\alpha'}|_{\partial(C_{\alpha\alpha'})}\cr
&&- \sum_{\alpha\alpha'\beta'} df^{\alpha\alpha'\beta'} - \sum_{\alpha\beta\alpha'}df^{\alpha\beta\alpha'}
\eea
and we wish to show that this is equal to 
\bea
\A^{\a}-\A^{\a'} = \sum_{\alpha,\alpha'} \int_{C_{\alpha\alpha'}} \(B^{\alpha} - B^{\alpha'}\) - \sum_{\alpha\beta} \Lambda^{\alpha\beta} + \sum_{\alpha'\beta'} \Lambda^{\alpha'\beta'}
\eea
Recalling that $B^{\alpha}-B^{\alpha'} = d\Lambda^{\alpha\alpha'}$, we see that the first term agrees. So it remains to understand that the other terms also agree. We may check that
\bea
&&\sum_{\alpha,\alpha'} \Lambda^{\alpha\alpha'}|_{\partial(C_{\alpha\alpha'})} - \sum_{\alpha\beta} \Lambda^{\alpha\beta} + \sum_{\alpha'\beta'} \Lambda^{\alpha'\beta'}\cr
&=& \sum_{\alpha,\alpha',\beta'} \(\Lambda^{\alpha\alpha'}+\Lambda^{\alpha'\beta'}+\Lambda^{\beta'\alpha}\)\cr
&&+\sum_{\alpha,\beta,\alpha'} \(\Lambda^{\alpha\beta}+\Lambda^{\beta\alpha'}+\Lambda^{\alpha'\alpha}\)
\eea
where the evaluation of these expressions are at the intersection points $C^{\alpha}\cap C^{\alpha'}\cap C^{\beta'}=:C^{\alpha\alpha'\beta'}$ and $C^{\alpha\beta\alpha'}$ respectively. Noting the cocycle condition, these two terms can be written as
\bea
\sum_{\alpha\alpha'\beta'} df^{\alpha\alpha'\beta} + \sum_{\alpha\beta\alpha'}df^{\alpha\beta\alpha'}
\eea
which exactly cancel the remaining two terms in $d\Lambda^{\a\a'}$ and we are done.

Finally we find a cocycle condition on triple overlaps $\U^{\a\a'\a''}$,
\bea
\Lambda^{\a\a'}+\Lambda^{\a'\a''}+\Lambda^{\a''\a} \equiv 0 {\mbox{ mod $2\pi$}}.
\eea
To see that, we first note that 
\bea
&&\sum_{\alpha,\alpha'} \int_{C_{\alpha\alpha'}} \Lambda^{\alpha\alpha'} 
+\sum_{\alpha',\alpha''} \int_{C_{\alpha'\alpha''}} \Lambda^{\alpha'\alpha''} 
+\sum_{\alpha'',\alpha} \int_{C_{\alpha''\alpha}} \Lambda^{\alpha''\alpha} \cr
&=& \sum_{\alpha,\alpha'\alpha''} \int_{C_{\alpha\alpha'\alpha''}} \(\Lambda^{\alpha\alpha'}+\Lambda^{\alpha'\alpha''}+\Lambda^{\alpha''\alpha}\) \cr
&=& \sum_{\alpha,\alpha'\alpha''} f^{\alpha\alpha'\alpha''}
\eea
The remaining terms combine with this one and give us
\bea
(\delta\Lambda)^{\a\a'\a''} = \sum_{\alpha\beta\alpha'\alpha''} (\delta f)^{\alpha\beta\alpha'\alpha''} + \sum_{\alpha\alpha'\beta'\alpha''} (\delta f)^{\alpha\alpha'\beta'\alpha''} + \sum_{\alpha\alpha'\alpha''\beta''} (\delta f)^{\alpha\alpha'\alpha''\beta''}\label{ccoc}
\eea
from which the desired cocycle condition follows from Eq (\ref{coc}).

We have thus showed that 
\bea
\A(C):=\int ds A_{\mu s}(C)\delta C^{\mu}(s)
\eea
defined as the transgression of $B$, is a connection one-form on a line bundle with transition functions
\bea
g^{\a\b}(C)\equiv e^{i\Lambda^{\a\b}(C)}
\eea
defined on double overlaps $\U^{\a\b}$ of open charts in $LM$.

We have some freedom to choose the transition functions, which is the gauge freedom 
\bea
\Lambda^{\a\b} &\rightarrow& \Lambda^{\a\b} + \lambda^{\a} - \lambda^{\b}\cr
\A^{\a} &\rightarrow& \A^{\a} + d\lambda^{\a}.
\eea
The relation between this gauge parameter and the local ones is
\bea
\lambda^{\a} = \sum_{\alpha\in \a} \int_{C^{\alpha}} \lambda^{\alpha} - \sum_{\alpha,\beta\in \a} \int_{C^{\alpha\beta}} h^{\alpha\beta}
\eea
and one may check that these gauge transformations in $LM$ are equivalent with the corresponding gauge transformations in $M$ given through Eq's (\ref{gaugevariation}).

By using this gauge freedom, we can bring the gauge field in the gauge
\bea
\partial^{\mu}_s A^{\a}_{\mu t} = 0.
\eea
In terms of the local two-form gauge field, this gauge condition reads
\bea
\partial^{\mu}B_{\mu\nu} = 0.
\eea

The gauge field strength is given by
\bea
\F = d\A := \frac{1}{2}F_{\mu s,\nu t}(C) dC^{\mu s}\wedge dC^{\nu t}
\eea
Using Eq (\ref{transgress}) we find that
\bea
F_{\mu s,\nu t} = H_{\mu\nu\rho}(C(s)) \dot{C}^{\rho}(s) 2\pi \delta_t(s).
\eea

\subsection{The Wilson surface}
The Wilson line in $LM$ is given by 
\bea
W(\Sigma)&=& \exp i \int_{\Gamma} \A
\eea
where
\bea
\int_{\Gamma} \A := \sum_{\a} \int_{\Gamma^{\a}} \A^{\a} - \sum_{\Gamma^{\a\b}} \Lambda^{\a\b}
\eea
Here $\Gamma$ is a path in $LM$. If we view a surface $\Sigma\subset M$ as the total space of a fiber bundle $\pi:\Sigma\rightarrow \gamma$, then $\Gamma=\pi^{-1}(\gamma)$ is a path in $LM$ that corresponds to $\Sigma$. Of course there are many different fibrations of $\Sigma$. The Wilson line in $LM$ has to be the same for any such fibration in order to be a honest Wilson surface (in $M$). This follows from the fact that $\A$ is the transgression of $B$, which can be seen to imply that $\int_{\Gamma} \A = \int_{\Sigma} B$ with the appropriate interpretations of both sides.

Again we can express the change of the Wilson surface under a gauge variation as
\bea
W(\Sigma)&\rightarrow& g(C)W(\Sigma).
\eea
where $\partial \Sigma = C$ may be one or several disjoint boundary loops, and
\bea
g(C) := \exp i\int_C \lambda.
\eea
We see that the group element is subject to the condition
\bea
g(\Omega C) = g(C)^{-1}
\eea
where $\Omega$ reverses the orientation. Here the orientation of the boundary $C$ is that which is induced by the orientation of the surface $\Sigma$. So for instance if the Wilson surface is a cylinder, we have two disjoint boundaries $C=C_1 \cup \Omega C_2$ with induced orientations such that the Wilson surface transforms like 
\bea
W(\Sigma;C_1,C_2) &\rightarrow& g(C_1)W(\Sigma;C_1,C_2)g(\Omega C_2)\cr
&=& g(C_1)W(\Sigma;C_1,C_2)g(C_2)^{-1}.
\eea
which mimics the way the Wilson line transforms under gauge transformations. We also notice that $C_2$ (but not $\Omega C_2$) is homotopic to $C_1$.

\subsection{Magnetic charge}
Let $M_3$ be a three-cycle in $M$ and let us pick some fibration $\pi: M_3 \rightarrow M_2$. That is, locally, $M_3$ is a like a product $M_2\times C$ where the fiber $C$ is a loop and $M_2$ is a two-manifold. If $\A$ is a connection on loop space $LM$, then the pull-back $\pi^*\A$ of this connection one-form to $M_2$ (when evaluated on a loop in $LM$ which is a fiber in $M_3$) is still a connection one-form and hence the curvature of this pull-back one-form to $M_2$ defines an element $\pi^* \F \in H^2(M_2,\mb{Z})$. Moreover, the transition functions on $M_2$ are precisely the $\Lambda^{\a\b}(C)$ where we restrict the $C$'s to be fibers in the bundle (or the fibration) with total space $M_3$. Since these transition functions can be expressed in terms of $f^{\alpha\beta\gamma}$, one could suspect that
\bea 
\int_{M_2} \pi^*\F = \int_{M_3} H\label{eo}
\eea
where $H=dB$. It should be possible to show this equality using Eq (\ref{ccoc}). However it seems to be easier to show this equality without invoking the transition functions which are not really needed to show this type of equality. All we need to know is that the field strenghts are globally well-defined forms. We may then first show that 
\bea
\int_{C_X} H = \pi^*\F.\label{ep}
\eea
Here the left hand side is what is called `integration along the fiber' and is denoted as the `push-forward' $\pi_* H$ in \cite{Bott-Tu}, and is given by
\bea
\int_{C_x} H := \int \frac{ds}{2\pi} H_{\mu\nu\rho}(C_X(s))\dot{C}_X^{\rho}(s)\delta C_X^{\mu}(s)\wedge \delta C_X^{\nu}(s)
\eea
where $C_X := \pi^{-1} (X)$ is the fiber over the point $X\in M_2$. To establish Eq (\ref{ep}) we note that 
\bea
\F = \int \frac{ds}{2\pi} H_{\mu\nu\rho}(C(s))\dot{C}^{\rho}(s)\delta C^{\mu}(s)\wedge \delta C^{\nu}(s)
\eea
and $\pi^*\F$ is obtained by putting $C=\pi^{-1}(X)$ into this result, and we see that it is indeed equal to $\int_{C_X} H$. 

We then apply the projection formula (proposition 6.15 in \cite{Bott-Tu}). Adapted to this situation this formula reads
\bea
\int_{M_3} H = \int_{M_2} \pi_* H
\eea
and from this Eq (\ref{eo}) immediately follows.

\section{Non-abelian generalization}\label{nonabel}
From the loop space perspective, the natural non-abelian generalization of a gerbe appears to be to take the transition functions 
\bea
g^{\a\b}(C) = e^{i\Lambda^{\a\b}(C)}
\eea
on a double overlap $\U^{\a\b}$, to be elements in some non-abelian gauge group $G$. To be more precise, we will take them to be group elements in the infinite tensor product of gauge groups $\bigotimes_s G_s$ where each $G_s$ is a copy of $G$. Here $s$ parametrizes the loops. We let $t^{a}$ be the generators of the gauge group $G$, and obey the Lie algebra
\bea
[t^a,t^b]=C^{ab}{}_c t^c.\label{lie}
\eea
As usual we can have different coupling constants $g^s$ in each factor $G_s$ of this gauge group. We choose a convention where we absorb these coupling constants in the connection $\A=A_{\mu s}dC^{\mu s}$. 

On a triple overlap $\U^{\a\b\c}$ the transition functions should be subject to the usual co-cycle conditions
\bea
g^{\b\c}(g^{\a\c})^{-1}g^{\a\b} = 1.
\eea

If the loops correspond to fundamental physical objects, then we should also demand that 
\bea
g(C)=g(C')
\eea
when $C'$ is a reparametrization of $C$ that is continuously connected with the identity map. We also require that
\bea
g(\Omega C) = g(C)^{-1}
\eea 
where $\Omega$ denotes orientation reversal of the loop.\footnote{These two conditions are nothing but special cases of the more general requirement that $g(C^n) = g(C)^n$ where $C^n$ is a loop with degree $n$ that is geometrically identical with $C$.} For this to become consistent with gauge transformations we must also require that
\bea
\A(\Omega C) = -\A(C).
\eea

\subsection{The Wilson surface}
We define the non-abelian Wilson surface $W(\Sigma)$ as an ordinary Wilson line in $LM$, 
\bea
W(\Gamma) = W(\Gamma^{\a}) g^{\a\b} W(\Gamma^{\b}) g^{\b\c} W(\Gamma^{\c}) \cdots
\eea
where $\Gamma$ is a fibration of $\Sigma$ and $\Gamma^{\a}\subset \U^{\a}$ are maximally contracted curve pieces, i.e. such that any non-empty overlap $\Gamma_{\a}\cap\Gamma_{\b}$ is a point in $LM$. Here
\bea
W(\Gamma^{\a}) = P \exp i\int_{\Gamma^{\a}} \A^{\a}
\eea
is defined by means of the differential equation
\bea
\frac{d}{dt} W = i\frac{d{C}^{\mu s}(t)}{dt}A^{\a}_{\mu s}W
\eea
of parallel transportation along $\Gamma^{\a}$ where we parametrize $\Gamma$ by $t\in [0,1]$ say. Then boundary loops $C_1$ and $C_2$ at $t=0$ and $t=1$ respectively of the curve $\Gamma^{\a}$ are defined such that they have orientations such that $C_1$ has the orientation that is induced by the orientation of the surface associated with $\Gamma^{\a}$ and $C_2$ has the orientation that makes it homotopic with $C_1$.

On any double overlap $\U^{\a\b}$ we have
\bea
\A^{\a} = g^{\a\b}\A^{\b}(g^{\a\b})^{-1} + ig^{\a\b} d (g^{\a\b})^{-1}.
\eea
Under a gauge transformation (change of trivialization), we have
\bea
g^{\a\b} \rightarrow  g^{\a} g^{\a\b} (g^{\b})^{-1}
\eea
for some group elements $g^{\a}$ defined over $\U^{\a}$. In order for the closed Wilson surface (which we define as $\tr W(\Gamma)$ where $\Gamma$ is a closed path in $LM$), to be gauge invariant, we must take 
\bea
W(\Gamma^{\a};C_1,C_2) \rightarrow g^{\a}(C_1)W(\Gamma^{\a};C_1,C_2)g^{\a}(C_2)^{-1}
\eea
which means that the connection one-form must transform as
\bea
\A^{\a} \rightarrow g^{\a}\A^{\a} (g^{\a})^{-1}+ig^{\a}d(g^{\a})^{-1}
\eea
which is nothing but the condition for the gauge covariant derivative to commute with gauge variations.

\subsection{Reparametrization invariance}
From the equation of parallel transportation we read off the gauge covariant exterior derivative $D=d-i\A$. The field strength may now be computed as the curvature of the connection,
\bea
\F &=& iD\wedge D\cr
&=& d\A - i\A\wedge\A.
\eea
It is locally a two-form that we can write as
\bea
\F = \frac{1}{2}F_{\mu s,\nu t} dC^{\mu s}\wedge dC^{\nu t}
\eea
It transforms covariantly under the gauge group,
\bea
\F \rightarrow g\F g^{-1}.
\eea
Infinitesimally this is
\bea
\F \rightarrow -i[\F,\Lambda]
\eea
For this to be a well-defined variation in the sense that the right-hand side be of the same form as the $\F$ we stared with, we must assume that
\bea
F_{\mu s,\nu s'} = 0 {\mbox{ whenever $s\neq s'$.}}\label{sam}
\eea

We can also define the field strength as a functional derivative of the Wilson line in $LM$ (in complete analogy with \cite{Polyakov}),
\bea
\frac{\delta W(0,1)}{\delta C^{\mu s}(t)} &=& W(0,t)F_{\mu s,\nu s'}(C(t))\dot C^{\nu s'}(t)W(t,1)\cr
&:=& \int ds' W(0,t)F_{\mu s,\nu s'}(C(t))W(t,1)\partial_t C^{\nu}(s',t).
\eea
where thus $t\mapsto C^{\mu s}(t) := ds C^{\mu}(s,t)$ is the path in $LM$ and $\dot{C}^{\mu s}(t):=\partial_t C^{\mu s}(t)$. 

Reparametrization invariance means that variations tangential to the surface vanish,\footnote{I am grateful to Urs Schreiber for having pointed this out to me. Another derivation can be found in \cite{Teitelboim,Gustavsson1,Gustavsson2}.}
\bea
\frac{\delta W}{\delta C^{\mu s}(t)} \partial_t C^{\mu}(s,t) &=& 0\cr
\frac{\delta W}{\delta C^{\mu s}(t)} \partial_s C^{\mu}(s,t) &=& 0
\eea
If we denote the pullback of the field strength to the surface $s^{\alpha}=(s,t)\mapsto C^{\mu}(s^{\alpha})$ as 
\bea
F_{\alpha s,\beta s'} = \partial_{\alpha}C^{\mu}(s,t)\partial_{\beta}C^{\nu}(s',t)F_{\mu s,\nu s'}
\eea
the we get the conditions for reparametrization invariance as
\bea
\int ds'F_{ts,ts'} &=& 0\label{ew1}\\
\int ds'F_{ss,ts'} &=& 0.\label{ew2}
\eea
Eq (\ref{ew1}) can be solved by requiring
\bea
F_{\mu s,\nu s'} = 0 \mbox{ if $s\neq s'$}\label{sol1}
\eea
and Eq (\ref{ew2}) is nothing but the constraint
\bea
\dot{C}^{\mu}(s)F_{\mu s,\nu s'} = 0.\label{sol2}
\eea
Now, how does this solution differ from the one given in for instance \cite{Schreiber}? One may check that Eq (\ref{sol2}) also holds for that field strength. However Eq (\ref{sol1}) does not. So, could there be another way of solving Eq (\ref{ew1}) where we do not require the kind of locality expressed by Eq (\ref{sol1})? Apparently there is one, and that is the one given in \cite{Schreiber} et. al. However that solution is not expressed in loop space. It relies on a certain flatness condition imposed on a connection one-form in space-time. If one wants to solve Eq (\ref{ew1}) in loop space, then there appears to be no other (covariant) way this can be done, but to assume that Eq (\ref{sol1}) holds.

Note that we thus again find the same condition Eq (\ref{sam}) as we found in a different way by demanding closure of gauge transformations and the assumption of a that the gauge group is an infinite tensor product (an assumption that this latter argument, based on reparametrization invariance, did not rely on).

\subsection{Loop algebra}
We will now explain how the multiplication is supposed to be carried out between $g(C)$ and $W(\Sigma)$.

If we to each boundary loop $C_i$ of the open Wilson surface, associate an index $s_i$,  
\bea
W(\Sigma;C_1,C_2,...)_{s_1 s_2...},
\eea 
then the gauge group should act on it as\footnote{Here we assume the induced orientation of all the boundary loops.}
\bea
W_{s_1s_2...}\rightarrow g_{s_1}{}^{s'_1}(C_1)g_{s_2}{}^{s'_2}(C_2)...W_{s'_1s'_2...}.
\eea
Then, if one boundary loop would be a pinched loop, let us say that it is $C_1\cup C_2=C$, then if we would interpret this as just one loop we would associate to it just one group element $g(C)$. If interpreted as two loops we would associate to it the group element $g(C_1)\otimes g(C_2)$. We should therefore require that
\bea
g(C_1\cup C_2)=g(C_1)\otimes g(C_2)
\eea
if we want a group element to be associated with each loop. If we then continuously merge and deform many small loops into one big loop $C$, it seems unlikely that this could spoil the tensor product property of the group element.\footnote{The tensor product would become more obvious if we associated group elements with open strings as well. Then we would compose two open strings by taking the tensor product of the associated group elements. Associating such group elements to open strings should not be confused with surface holonomies that can parallel transport open strings. In this paper group elements associated with closed surface holonomies are associated with the parallel transport of closed strings only, and should not be confused with the group elements $g(C)$.}

On the algebra level, an infinite tensor product of group elements should correspond to a loop algebra 
\bea
[t^a_s,t^b_t] = \delta_s(t) C^{abc} t^c_t.
\eea
If we work on a metric space, then the coupling constants $g^s$ (that in this form thus transform contravariantly under reparametrizations) can be related to the induced metric $g_{ss}$ on the loops. We find this to be convenient. But if we relate the $g^s$ to the metric, then we must later show that the metric-dependence is unphysical. We will relate the coupling constants to the metric as
\bea
g^s = \sqrt{g}g^{ss}.
\eea
In this expression, the index $s$ is to be interpreted as a vector index in a one-dimensional space of a loop, rather than as a continuous in an infinite-dimensional loop space.

We will normalize the generators as
\bea
\tr\(t^a_s t^b_t\) = \frac{1}{l(C)}\delta^{ab} \delta_{st}
\eea
where $\delta_{st} := \sqrt{g}\delta_s(t)$ and 
\bea
l(C) = \int ds \sqrt{g}
\eea
is the invariant length of the loop. We then find that
\bea
\tr\(t^a t^b\) = \delta^{ab}.
\eea
We also define
\bea
t^{an}_s := D^n t^a_s
\eea
Here $D^n$ is the metric covariant derivative rised to the power $n$ in such a way that it transforms as a scalar, i.e.
\bea
D^{2m} &:=& (D^s D_s)^m\cr
D^{2m+1} &:=& \sqrt{g}g^{ss}D_s D^{2m}.
\eea

The most general form of the gauge parameter is 
\bea
\lambda(C) = \int ds \sqrt{g}g^{ss}\lambda^{an}_s t^{an}_s
\eea
though we could perform integrations by parts and get an expression where we have only the generators $t^a_s$. 

The gauge field should be of the same form as the gauge parameter. Then covariance dictates it to be of the form 
\bea
\A(C) = \int ds \sqrt{g}g^{ss} A^{an}_{\mu s}(C)\delta C^{\mu}(s)t^{an}_s.\label{standardgauge}
\eea
As we will see below, the metric-dependence of the connection can be gauged away. 
Given the form of the gauge parameter, it is clear that the gauge field should be of this form if it is pure gauge. Another justification for this form comes from the requirement that the Wilson surface be reparametrization invariant. More precisely, from Eq (\ref{sol1}).

\subsection{Gauge transformations revisited}
Infinitesimally a gauge variation of the gauge field is given by
\bea
\delta_{\lambda} \A = D\lambda
\eea
where
\bea
D=d-i\A
\eea
is the gauge covariant exterior derivative, and let us assume that we have brought the gauge parameter into the form
\bea
\lambda = \int ds \sqrt{g}g^{ss}\lambda_s^a t^a_s.
\eea
As usual we now find that 
\bea
[\delta_{\mu},\delta_{\lambda}] \A = D\nu
\eea
where 
\bea
\nu = -i[\mu,\lambda].
\eea
We must now insure that the new gauge parameter $\nu$ is of the same form as the original ones. We find that
\bea
\nu = \int ds \sqrt{g}g^ss \nu_s^a t^a_s
\eea
with\footnote{The `i' is just due to the fact that we have chosen a convention where we have purely imaginary structure constants.}
\bea
\nu^a_s = -i\sqrt{g}g^{ss}C^{abc}\mu^b_s\lambda^c_s.\label{nu}
\eea

In order for the gauge variation of $\A$ to be of the same form as $\A$, we must require a locality condition
\bea
D_{\mu s}\lambda_t = 0{\mbox{ if $s\neq t$}}.
\eea
where we may define $\lambda_t := \sqrt{g}g^{tt}\lambda_t^a t^a_t$. If now $\lambda_s$ and $\mu_s$ are subject to such locality conditions, then we find that also $\nu_s$ is subject to this same locality condition, by applying the Leibniz rule for differentiation on Eq (\ref{nu}).

We can express the locality condition as\footnote{A third way of expressing this would be by saying that $\lambda_s(C)$ must depend only locally on $C$ via $D^n C(s)$ for $n=0,1,2,...$.}'\footnote{This locality condition would not close to under a gauge algebra if we would truncate the series and only let the gauge parameter depend on say $C(s)$ and $\dot{C}(s)$. We have to include the whole set of derivatives $D^n C(s)$ for all $n$ to get a closed gauge algebra.}
\bea
D_{\mu s}\lambda_t^a = \sum_{n=0}^{\infty} \xi^{an}_{\mu s}(C) D^n\delta(t-s)
\eea
for some vector fields $\xi_{\mu s}^{an}$'s. We now find that
\bea
\delta A_{\mu s}^{an} = \xi^{an}_{\mu s}
\eea
though this expression is not very illuminating, so let's define
\bea
A_{\mu s}:= \sqrt{g}g^{ss}A_{\mu s}^a t^{an}_s
\eea
so that
\bea
\A = \int ds \delta C^{\mu}(s) A_{\mu s}.
\eea
Then we find that
\bea
\delta A_{\mu s} = D_{\mu s}\lambda.
\eea

We can pick out the component $\delta A_{\mu s}^{an}$ by taking the trace with some $t^{bm}$, using that
\bea
\tr\(t^{an}_s t^{bm}_t\) = \frac{1}{l(C)}\delta^{ab}D^{n}(s)D^m(t) \delta_{st}.
\eea
But now we would like to take $m=-n$ to isolate the delta function on the right-hand side. We can indeed extend 
\bea
t^{an}_s = D^n t^a_s
\eea
to any complex number $n$ by analytic continuation.\footnote{This can be done by covariantizing the definition of the fractional derivative, and thus define
\bea
D^n f(t) := \frac{1}{\Gamma(-n)} \int_t^{\infty} ds \sqrt{g}\frac{f(s)}{(t-s)^n}
\eea
where $\Gamma(n+1)=n\Gamma(n)$ denotes the usual Gamma function.}

It now seems plausible that we can always make a gauge transformation that brings the gauge potential into Lorentz gauge
\bea
\partial^{\mu}_t A_{\mu s} = 0
\eea
Upon a gauge variation of this we find in part the operator $\partial^{\mu}_t \partial_{\mu s}<0$ which is invertible when acting on the gauge parameter that depends only locally on a point on the loop (which in effect means that we get non-zero eigenvalues of $\partial^{\mu}_t\partial_{\mu s}$ only on the diagonal $t=s$ where this operator is negative). But now we should rather consider $\partial^{\mu}_t D_{\mu s}$. In Yang-Mills theory one may argue that this operator must also be invertible as this to zeroth order in the coupling constant coincides with $\partial^{\mu}_t \partial_{\mu s}$. In the application to $(2,0)$-theory where we have no adjustable coupling constant, we do not know how to prove that one can always impose Lorentz gauge, but we will we assume that this is the case also here.

\subsection{Metric-independence}
We will now show that the metric-dependence of the connection can always be gauged away. 

Let us consider a surface parametrized by $s$ and $t$, embedded in $M$ as $(s,t)\mapsto X^{\mu}(s,t)$, and take the pullback of the gauge field one-form to this surface as
\bea
dt \int ds A_{ts} = dt\int ds \frac{\partial X^{\mu}(s,t)}{\partial t} A_{\mu s}(C_t)
\eea
where $C_t^{\mu}(s)=X^{\mu}(s,t)$ is the constant $t$ loop in the surface. Since $A_{\mu s}$ depends on $C$ only via $D^n C(s)$, it follows that the pullback $A_{ts}$ depends locally on $s^{\alpha}:=(s,t)$. Furthermore, in order to be able to integrate this over the surface, it must transform like a two-form, i.e. $dt\wedge ds A_{ts} = \frac{1}{2}ds^{\alpha}\wedge ds^{\beta} A_{\alpha\beta}$ is a two-form on the surface.\footnote{We think it is desirable to be able to integrate this pullback over the surface as that makes it possible to express the Wilson surface as a Dyson series expansion.}

As before, we assume that one can always make a gauge transformation (infinitesimally of the form $\delta A_{\alpha s} = D_{\alpha s}\Lambda$ when pulled back to the surface), that brings the gauge field into the Lorentz gauge
\bea
D^{\alpha} A^a_{\alpha s} = 0
\eea
where $D_{\alpha}$ is the metric compatible derivative. Using that $A_{\alpha\beta}$ is a two-form, we can rewrite this condition in the form (the labels $s$ and $t$ may now be freely interchanged since $A_{\alpha\beta}$ depends locally on them and we could just as well have started with a constant $s$ loop $C$ being parametrized by $t$ instead of $s$ to arrive at the condition $D^{\alpha} A_{\alpha t}^{a} = 0$)
\bea
D^{s} A^a_{ts} = 0\label{lorentz}
\eea
and this is precisely what we need to be able to write the connection one-form in the manifestly metric-independent form
\bea
\A = \int ds A^a_{\mu s}\delta C^{\mu}(s) t^a.\label{cova}
\eea
Let us show this in some more detail. Apriori the gauge field is given by
\bea
\A = \int ds \sqrt{g}g^{ss}A_{\mu s}^{a}\delta C^{\mu}(s)D^n t^a_s.
\eea
Making integrations by parts, we get (up to a sign)
\bea
\A \sim \int ds D^n \(\sqrt{g}g^{ss}A_{\mu s}^{a}\delta C^{\mu}(s)\) t^a_s.
\eea
All terms with $n\geq 1$ vanishes in Lorentz gauge by Eq (\ref{lorentz}), and by the fact that the two-dimensional surface parametrized by $s$ and $t$ really was chosen completely arbitrarily. So all what remains in Lorentz gauge is the term with $n=0$, 
\bea
\A = \int ds \sqrt{g}g^{ss}A_{\mu s}^{a}\delta C^{\mu}(s) t^a_s.
\eea
This term must be equal to 
\bea
\A = \int ds A_{\mu s}^{a}\delta C^{\mu}(s) t^a.
\eea 
To understand that, we first note that both these expressions are invariant under reparametrizations of $s$. So if we show that they are equal for one parametrization, they are equal for any parametrization. Second, we note that 
\bea
\partial_s \(\sqrt{g}g^{ss}A_{\mu s}^{a}\delta C^{\mu}(s) \) = 0
\eea
by Eq (\ref{lorentz}) and hence the integrand is really just over $t^a_s$. Let us then fix the parametrization by making the gauge choice $g_{ss} = 1$. Then we get
\bea
\A = A_{\mu s}^{a}\delta C^{\mu}(s) t^a
\eea
where $s$ can be chosen to be any value in the interval $[0,2\pi]$ since this in any case is just a constant with all these gauge choices being made. At no cost at all we can hang on an integral sign so as to end up with Eq (\ref{cova}). The virtue with hanging on the integral sign is of course that we then get an expression that is valid in any parametrization, and not just in the gauge $g_{ss}=1$.

The alert reader might have wondered if we really have showed the metric independence now, since we must use the metric to formulate the Lorentz gauge condition. 

Let us assume that we have a fibration of the space-time $M$ and of the Wilson surface $\Sigma$, such the projector $\pi$ associated with the total bundle space $\Sigma$ is the restriction of the projector associated with space-time $M$,
\bea
\pi:&&M\rightarrow M_5\cr
\pi:&&\Sigma\rightarrow \gamma
\eea
Let us denote coordinates in $M_5$ as $X^I$ and fibers in $M$ by $C_X = \pi^{-1}(X)$. Restricting $X$ to the surface $\Sigma$ we thus get fibers that lie in $\Sigma$ by our assumption. The connection on $LM$ gets projected to a connection on $M_5$ via the pullback map 
\bea
\pi^*: A_{\mu s}(C_X) \mapsto A^{\pi}_I(X) = A_{\mu s}(C_X)\frac{\partial C^{\mu s}_X}{\partial X^I}
\eea
and we deduce that, at least in Lorentz gauge, the closed Wilson surface can be expressed exactly as a Wilson loop in $M_5$
\bea
\tr P \exp i \int_{\gamma} A^{\pi}.
\eea

Let us summarize the chain of steps we have taken to reach this conclusion: We could only make the identification between the Wilson surface and the Wilson loop explicit in Lorentz gauge. But since both the closed Wilson surface and the Wilson loop are gauge invariant objects, they must agree for any gauges. Finally we noted that the Wilson loop is metric independent and concluded that so must also the closed Wilson surface be.

\subsection{Magnetic charge}\label{open}
Generalizing the concept of the abelian magetic charge, we would like to define the non-abelian magnetic charge vector associated with a three-manifold $M_3\subset M$ as
\bea
\int_{M_2} \pi^*\F
\eea
where $M_2$ is the base manifold in any fibration of $M_3$ with projector $\pi: M_3 \rightarrow M_2$. If we go to Lorentz gauge we find that 
\bea
\F = \int ds \int ds' F^A_{\mu s,\nu s'}(C)\delta C^{\mu}(s)\wedge \delta C^{\nu}(s')t^A
\eea
To get $\pi^*{\F}$, we evaluate $\F$ on loops that are fibers in the fiber bundle $M_3$, and take the pullback to $M_2$. We assume a maximally broken gauge group, i.e a product of $U(1)$ factors, and we may always assume that we have gauge rotated $\F$ so as to lie in the Cartan subalgebra of the original Lie algebra associated with the gauge group $G$. The $t^A$ generators are thus the Cartan $U(1)$ generators in the broken gauge group.

We now have to establish that this definition is independent of how we fibrate $M_3$. This follows if one can show that the pullback field strength $\pi^* \F$ defines an element in the first Chern class $c_1(\F) = H^2\(M_2,2\pi{\mb{Z}}\)$. Then a continuous deformation of the fibration can not change this discrete topological class. That $\pi^* \F$ defines an element is $c_1(\F)$ follows if we notice that the pullback of the gauge connection, $\pi^* \A$ to $M_2$ is really a connection on $M_2$ with transition functions $g^{\a\b}(\pi^{-1}(X))$ defined as the pullback of the transition functions on $LM$ and open charts on $M_2$ are pullbacks of open charts on $LM$,
\bea
U^{\a} := \pi^* \U^{\a} := \{X\in M_2 | \pi^{-1}(X) \in \U^{\a}\}.
\eea

\subsection{Ultra-local expressions?}
One may wonder if we really need to consider these subtle fields on loop space which depend on all derivatives $D^n C(s)$ of the loop. So let us make the following ultra-local ansatz
\bea
\A(C) &=& \int \frac{ds}{2\pi} \sqrt{g}g^{ss}B^{an}_{\mu\nu}(C(s))\dot{C}^{\nu}(s)\delta C^{\mu}(s)t^{an}_s\cr
\lambda(C) &=& \int \frac{ds}{2\pi} \sqrt{g}g^{ss}\lambda_{\mu}^{an}(C(s))\dot{C}^{\mu}(s) t^{an}_s
\eea
for the connection and gauge parameter. In this case we would find metric independent expressions 
\bea
\A(C) &=& \int \frac{ds}{2\pi} B^{a0}_{\mu\nu}(C(s))\dot{C}^{\nu}(s)\delta C^{\mu}(s)t^a\cr
\lambda(C) &=& \int \frac{ds}{2\pi} \lambda_{\mu}^{a0}(C(s))\dot{C}^{\mu}(s) t^a
\eea
if we had
\bea
\partial_s\(\sqrt{g}g^{ss}B^{an}_{\mu\nu}(C(s))\dot{C}^{\nu}(s)\delta C^{\mu}(s)\) &=& 0\cr
\partial_s\(\sqrt{g}g^{ss}\lambda^{an}_{\mu}(C(s))\dot{C}^{\mu}(s)\) &=& 0
\eea
If we consider the pullback of $\A$ to a two-manifold $\Sigma \subset M$ in which $C$ is embedded, and which we parametrize by $s$ and $t$, and let $\delta C^{\mu}(s) = dt \partial_t C^{\mu}(s,t) + ds \partial_s C^{\mu}(s,t)$ where $C^{\mu}(s)=C^{\mu}(s,t)$ for some constant $t$, then we can write these conditions in terms of the pullback fields as (noting the anti-symmetry of $B_{\mu\nu}$)
\bea
D^s B^{an}_{st} &=& 0\cr
D^s \lambda^{an}_s &=& 0
\eea
Such equations would follow as the pullback to the surface or loop, of the equations
\bea
\partial^{\mu}B_{\mu\nu}^a &=& 0\cr
\partial^{\mu}\lambda^a_{\mu} &=& 0
\eea
in flat six dimensions. Noting that 
\bea
A_{\mu s}^a(C) = B^a_{\mu\nu}(C(s))\dot{C}^{\nu}(s)
\eea
we can write the gauge condition as
\bea
\partial^{\mu}_t A_{\mu s}^a = 0.
\eea
Gauge transformation with gauge parameters that are subject to $\partial^{\mu}\lambda_{\mu}^a$ will shift the gauge field to another metric independent field configuration. But more general gauge parameters may bring in a metric dependence, but these are thus gauge equivalent with a metric independent configuration. 

Now a little experimentation quickly shows that the gauge algebra does not close on such ultra-local gauge fields. Making one gauge variation we get a non-local expression from the commutator term, that involves $\dot{C}^{\mu}(s)\dot{C}^{\nu}(t)$, although the non-locality is just an illusion that can be traded for a metric dependent connection, which is also just an illusion. Though we can not make both the metric-independence and the locality manifest simultaneosly. So we immediately find that we must at least extend the local expressions for $\A$ and $\lambda$ to include an arbitrary number of factors $\dot{C}(s)$. But not even that will close under the gauge algebra. When we compute $d\lambda$ of such a gauge parameter, we also find $\ddot{C}(s)$-terms and all higher derivatives. So both the gauge field and gauge parameter must depend on all the $D^nC(s)$.

Now, if we know all derivatives in one point $s$, then we can reconstruct the whole loop $C$. Does that mean that we have a non-local dependence on the loop then? I would say no. We had locality constraints of the form $\partial_{\mu s}A_{\nu t}=0$ if $s\neq t$. Let us take an example that could illustrate the point. Taylor expaning $f(x)= e^{-x^2}$ we get a power series $1-x^2+x^4/2+...$. If we now Fourier transform the Taylor series, with the argument $s-t$, then we get an infinite sum of delta functions $\delta(s-t) + \delta''(s-t) + \frac{1}{2}\delta''''(s-t) + ...$. Each finte partial sum in this series vanishes unless $s\neq t$, but the whole infinite sum should  behave very differently from this. Since if we Fourier transform $e^{-x^2}$ we get a smooth function that does not vanish for any $s-t$. This is how an infinte sum of delta functions should be interpreted as a smooth function, that in the situation at hand would correspond to a true non-local dependence on the loop. But that would thus violate the locality constraints. I would like to express this as saying that we have a local dependence on the loop at the point $s$, somehow. Expressing this by saying that we have a dependence on just a finite number of derivatives $D^n C(s)$ is insufficient. What then if we perform an infinite number of gauge transformations? Since the locality condition is obeyed for each gauge transformation, it should still be obeyed by after an infinite number of gauge transformations. But then we have a local dependence on $C$ at the point $s$ that depends on an infinte number of derivatives $D^n C(s)$. But this can not be a generic dependence on all derivatives at $s$ as that would mean a non-local dependence of $C$.

\section{$N=(2,0)$ supersymmetry}
In order to apply this formalism to $(2,0)$ theory, one would like to extend loop space to a super loop space. Translations in loop space are generated by $P_{\mu s}=-i\partial_{\mu s}$. This is not a rigid translation of the loops in space-time but allows for arbitrary deformations of the loops. The point is that $P_{\mu s}$ generate rigid translations in loop space, not in space-time. Supercharges should square to this generator of rigid translations. The supersymmetry algebra we would like to propose in loop space would therefore be
\bea
\{Q_s,Q_t\} = -2\delta_t(s)\Gamma^{\mu}P_{\mu s}.
\eea
We thus let the supercharges depend on the parameter $s$, though not on the point in loop space. These $Q_s$ should be generators rigid supersymmetry in loop space. Apriori the supersymmery parameter $\epsilon(s)$ could also be some arbitrary (Grassmann odd) function of $s$. Though we have not been able to find any supersymmetry multiplet that would respect such a big supersymmetry. Instead we restrict to (covariantly) constant parameters $\epsilon(s)=\epsilon$. Supersymmetry variations are then generated by 
\bea
\delta_{\epsilon s} := \epsilon Q_s.
\eea
The on-shell supersymmetry variations for the components fields were obtained in \cite{Gustavsson2}. It would be nice to see how to express this in terms of a superfield in a super loop space.

Eventually one would of course like to `quantize' the theory. However that does not seem to be so easily done. This is difficult because selfuality constrains the coupling constant to be a fixed `selfdual' number (of order one), that can never be made small. So even if we have a supersymmetric non-abelian classical action, we will not be able to use it to quantize the theory, at least not in a perturbative framework. Nevertheless a classical action can be useful for other purposes. It can be used to obtain classical solitonic solutions, and to study quantum theory for zero modes about such solutions.

\end{document}